\documentclass[11pt]{article}
\begin{document}

\def\half{{\frac{1}{2}}}
\def\r{\hat\rho}
\def\tr{{\rm tr}}
\def\s{\sigma}
\def\p{\hat p}\def\x{\hat x}
\def\C{{\bf C}}

%
\title{Wigner Centennial: His Function, and Its Environmental Decoherence} 
\author{Lajos Di\'osi\\
             Research Institute for Particle and Nuclear Physics\\
             H-1525 BUdapest 114, POB 49, Hungary}
\maketitle
\abstract{
In 1983, Wigner outlined a modified Schr\"odinger---von-Neumann equation
of motion for macroobjects, to describe their typical coupling to the
environment. This equation has become a principal model of environmental
decoherence which is beleived responsible for the emergence of
classicality in macroscopic quantum systems. Typically, this happens
gradually and asymptotically after a certain characteristic decoherence
time. For the Wigner-function, however, one can prove that it evolves
perfectly into a classical (non-negative) phase space distribution
after a finite time of decoherence.}

 

\section{Introduction}\label{intro}
Wigner, in his paper {\it On the Quantum Correction For Thermodynamic 
Equilibrium} \cite{Wig32}, constructs a map from the {\it quantum} state of 
a particle into a certain {\it classical} phase space distribution:
\begin{equation}\label{rtoW}
\r \longrightarrow W(x,p)~.
\end{equation}
(Wigner adds in footnote that it  was found `by. L.Szilard and the 
present author some years ago for another purpose.') 
The Wigner-function $W(x,p)$ became the prototype of all subsequent
trials to simulate a quantum state through a statistical distribution over
the classical phase space. 
As it was already obvious to Wigner himself, the sign of $W(x,p)$ is 
indefinite. By now, it is widely accepted that the 
existence of domains where $W$ is negative should mean that the state $\r$ 
is essentially quantum in a sense that it can not be simulated by a 
statistical distribution $W(x,p)$ over the classical phase space. 
In the contrary case, when $\r$'s Wigner-function is positive, 
one can say that $\r$ exhibits classicality in the above particular sense.

After half century \cite{Wig83}, Wigner was certainly the first among the 
most influential to support the concept later called environmental 
decoherence \cite{Zur91,deco}. Wigner, according to his 
{\it Review of the quantum mechanical measurement problem},
was impressed \cite{Wigcit} by the work of Zeh \cite{Zeh71} claiming that a 
macroscopic body can actually not be a {\it closed} system of its microscopic 
degrees of freedom. Wigner adopts such a reality and emphasizes the
need of a {\it new equation} for apparently non-isolated objects.
He modifies the Schr\"odinger equation by adding a second term to the r.h.s.:
\begin{equation}\label{Schmod}
\frac{d\r}{dt}=-\frac{i}{\hbar}[{\hat H},\r]+ {\rm irreversible~term}~.
\end{equation} 
The irreversible term models the inevitable interactions
of the macroscopic body with its environment. Perhaps, this was the
first closed equation with the explicite intention of modeling 
environmental decoherence.

Decoherence \cite{Zur91,deco} is responsible for the emergent classicality in
quantized systems. In particular, the negative valued domains of the
Wigner-function $W(x,p)$ might become washed out and grow positive
under the influence of a modified Schr\"odinger equation 
like Eq.~(\ref{Schmod}). Kiefer and the present author have recently
proved an even stronger theorem \cite{DioKie02}. 
The positivity of $W(x,p)$ is achieved in finite time, contrary to
our intuition that continuous variables would only asymptotically 
decohere. 

Sec.~\ref{HisFE} outlines the function and the equation
proposed by Wigner in 1932 and 1983, respectively. In Sec.~\ref{decoherence} 
the theorem of exact decoherence is recapitulated. 
 
\section{His Function (1932) and Equation (1983)}\label{HisFE}  
According to Wigner \cite{Wig32}, the quantum state $\r$ can be mapped
into a normalized phase space distribution:
\begin{equation}\label{W}
W(x,p)=\frac{1}{2\pi}\int
\langle x-r/2\vert\r\vert x+r/2\rangle
{\rm e}^{ipr/\hbar}{\rm d}r~.
\end{equation}
There is an equivalent form of the above map, see e.g. \cite{Dio96}:
\begin{equation}\label{W1}
W(x,p)=\tr\left[\r{\cal S}\delta(x-\x)\delta(p-\p)\right]
\end{equation}
where $\cal S$ stands for total symmetrization defined by successive
application of the rules
\begin{equation}\label{symm}
{\cal S}\x\hat f=\half\left(\x\hat f+\hat f\x\right)~,~~~~
{\cal S}\p\hat f=\half\left(\p\hat f+\hat f\p\right)~,
\end{equation}
to the already ${\cal S}$-ordered function $\hat f=f(\x,\p)$.
The form (\ref{W1}) shows directly that the Wigner-function has the
correct quantum mechanical marginal distributions for both canonical
variables separately:
\begin{eqnarray}\label{marg}
\int W(x,p){\rm d}p &=&\tr[\r\delta(x-\x)]~,\\
\int W(x,p){\rm d}x &=&\tr[\r\delta(p-\p)]~.
\end{eqnarray}
The form (\ref{W1}) of the Wigner-function is covariant against {\it linear} 
canonical transformations:
\begin{equation}\label{cancov}
\left({\x\atop\p}\right)\rightarrow\left({\x'\atop\p'}\right)~~\Rightarrow~~
W(x,p)\rightarrow W'(x',p')=W(x,p)~.
\end{equation}

Wigner proposes the following modification of Schr\"odinger's equation
in case of a macroscopic body \cite{Wig83}:
\begin{equation}\label{Weq}
\frac{d\r}{dt}=-\frac{i}{\hbar}[{\hat H},\r]
               -\sum_{\ell m}\varepsilon_{\ell}
                     \left[\hat L_{\ell m},[\hat L_{\ell m},\r]\right]~,
\end{equation}
where ${\hat H}$ is the total Hamiltonian and $\hat L_{\ell m}$ are
the multipole operators of the angular momentum. The second term
is intented to decohere macroscopically different {\it multipole moments}.
The strengths of their decoherence are being controlled by the rotation
invariant parameters $\varepsilon_{\ell}$. It turns out, however, that 
more typical is the decoherence between macroscopically 
different center of mass {\it positions} $\x$. The modified
Schr\"odinger equation retains the mathematical structure of Wigner's one 
(\ref{Weq}). Indeed, for a free object of mass $m$ it is this simple
\cite{JooZeh85}:  
\begin{equation}\label{xx}
\frac{d\r}{dt}
                 =-\frac{\rm i}{\hbar}\left[\frac{\p^2}{2m},\r\right]-
               \frac{D}{2\hbar^2}\left[\x,[\x,\r]\right]~.
\end{equation}
This equation simplifies the environment as if it corresponded
to a random external force $F(t)=\sqrt{D}w(t)$ where $w(t)$ is
standard white-noise. The solution of the standard Schr\"odinger
equation with the randomly fluctuating potential $F(t)\x$, averaged
over the noise $w$, yields the above modified Schr\"odinger 
equation \cite{Dio85}. 

The strength of decoherence is governed by a single parameter $D$.
Assume a certain coherent width $\s$ for the initial quantum
state. Dimensional analysis of the modified Schr\"odinger equation
shows that the Hamiltonian and the irreversible terms have opposite
tendencies. The unitary term increases $\s$ at a
`coherent broadening' characteristic time $m\sigma^2\hbar$.
The irreversible term is tending to decrease $\s$ at a time scale
$\hbar^2/D\sigma^2$ of `incoherent localization'. The two contrary
effects may have a balance at the `stationary coherence width' \cite{Dio87}
\begin{equation}\label{s0}
\sigma_0 = \left(\frac{\hbar^3}{Dm}\right)^{1/4}~,
\end{equation} 
achieved typically after a characteristic `decoherence time'
\begin{equation}\label{t0}
t_0=\sqrt{\hbar m/D}~.
\end{equation}
Then the physical picture suggests that the quantum state has become a
random (incoherent) mixture of wave packets whose characteristic
width is the stationary value (\ref{s0}). The corresponding
Wigner-function must obviously be positive since each contributing
wave packet is assumed to have a positive Wigner-function.

\section{Exact decoherence}\label{decoherence}
Let us consider the evolution of the quantum state $\r(t)$ under the
influence of environmental decoherence as described by Eq.~(\ref{xx}).
The following theorem holds \cite{DioKie02}. Independently of the initial 
state $\r(0)$, the solution $\r(t)$ will exhibit exact classicality for
$t \geq t_D$ in a sense that the corresponding Wigner-function becomes
non-negative:
\begin{equation}\label{theo}
W(x,p;t)\geq 0~~~~{\rm iff}~~t\geq t_D~.
\end{equation}
The exact value of the decoherence time is calculable: $t_D=3^{1/4}t_0$.

The proof is the following. Substituting Eq.~(\ref{W}) into Eq.~(\ref{xx}),
we obtain the standard classical Fokker-Planck equation for the 
Wigner-function 
(\ref{W}): 
\begin{equation}\label{FP}
\frac{dW}{dt} = -\frac{p}{m}\frac{\partial W}{\partial x} 
                 +\frac{D}{2}\frac{\partial^2 W}{\partial p^2}\ .
\end{equation}
Consider the Gaussian of width $\sqrt{C}$ in both $x$ and $p$:
\begin{equation}\label{gC}
g(x,p;C)= \frac{1}{2\pi C}\exp
\left[-\frac{x^2+p^2}{2C}\right]~,
\end{equation} 
and use it to coarse-grain an arbitrarily chosen Wigner-function. The result 
is always non-negative if and only if
the coarse-graining scale is greater than one-half: 
\begin{equation}\label{Husimi}
g(x,p;C)\star W(x,p)\geq0~~~~{\rm iff}~~C\geq1/2~,
\end{equation}
where $\star$ denotes convolution.
Indeed, for $C=1/2$ the coarse-graining yields the positive Husimi-function 
\cite{Per93}; greater values of $C$ yield further coarse-graining 
which preserves the positivity of the Husimi-function. 
Now we can generalize the lemma (\ref{Husimi}).
Let us generalize the Gaussian profile (\ref{gC}) first:
\begin{equation}\label{gC1}
g(x,p;\C) = \frac{1}{2\pi}\vert\C\vert^{-\half}
          \exp\left[-(x,p)\frac{1}{2\C}\left({x\atop p}\right)\right]~.
\end{equation}
The following lemma holds, see also Ref.~\cite{KhaTsi92} by Khalfin and 
Tsirelson (1992):
\begin{equation}\label{lemma}
g(x,p;\C)\star W(x,p) \geq0~~~~iff~~~~\vert\C\vert\geq1/4~.
\end{equation}
The correlation matrix $\C$ can always be transformed into the form 
$C{\bf I}$ with $C=\sqrt{\vert\C\vert}$.
Hence, due to the covariance (\ref{cancov}) of the Wigner-function,
the lemma (\ref{lemma}) follows from the special case (\ref{Husimi}).

Coming back to the Fokker-Planck equation (\ref{FP}), its solution can
be written as the following time-dependent Gaussian coarse-graining:
\begin{equation}\label{FPsol}
W(x,p;t)=g(x,p;\C_W(t))\star W(x-pt/m,p;0)~.
\end{equation}
The correlation matrix of the coarse-graining profile is time-dependent: 
\begin{equation}\label{Ct}
\C_W(t)=
Dt\left(\begin{array}{cc}t^2/3m^2&t/2m\\t/2m&1\end{array}\right)~,
\end{equation}
as it can be inspected if we insert it into Eq.~(\ref{FPsol}) which
we substitute into (\ref{FP}). The determinant yields:
\begin{equation}\label{det}
\vert\C_W(t)\vert=\frac{D^2t^4}{12m^2}~.
\end{equation}
The proof of the theorem (\ref{theo}) culminates in the observation that 
the determinant is monotone function, i.e., the Wigner-function $W(x,p;t)$ 
results from progressive coarse-graining. Applying the lemma (\ref{lemma}) 
to the determinant, we conclude that the Wigner-function $W(x,p;t)$ is 
non-negative if $\vert\C_W(t)\vert\geq1/4$. This condition is equivalent with
$t\geq3^{1/4}t_0\equiv t_D$. This completes the exact proof of the
theorem whose earlier versions can be found in Refs.~\cite{earlier}.

\section{Conclusions}\label{concl}
I discussed two items among Wigner's remarkable contributions to the 
foundations of quantum mechanics: the first phase-space quasi-distribution 
(1932) and the first explicite equation of decoherence (1983), respectively. 
The simplified version of his decoherence equation leads to a theorem when
applied to his function $W(x,p;t)$. I recapitulated the exact proof,
given recently by Claus Kiefer and myself, guaranteeing the positivity of 
Wigner's function after a finite time of decoherence. The purpose of my talk
was to demonstrate Wigner's continued impact even on the second century of 
quantum mechanics.

\section*{Acknowledgement(s)}
My research was supported by the Hungarian OTKA Grant No. 032640.
 

\end{document}